\documentclass{aa}  
\usepackage{graphicx}
\usepackage{txfonts}
\usepackage{natbib}

\bibpunct[]{(}{)}{;}{a}{}{,}
\def\ltsima{$\; \buildrel < \over \sim \;$}
\def\simlt{\lower.5ex\hbox{\ltsima}}
\def\gtsima{$\; \buildrel > \over \sim \;$}
\def\simgt{\lower.5ex\hbox{\gtsima}}
\begin{document}
   \title{Strongly induced collapse\\ in the Class 0 protostar NGC~1333 IRAS~4A}


   \author{A. Belloche
          \inst{1}
          \and
          P. Hennebelle
          \inst{2}
          \and
          P. Andr\'e\inst{3,4}
          }

   \offprints{A. Belloche}

   \institute{Max-Planck-Institut f\"ur Radioastronomie, Auf dem H\"ugel 69,
              D-53121 Bonn, Germany\\
              \email{belloche@mpifr-bonn.mpg.de}
         \and
              LERMA/LRA, Ecole Normale Sup\'erieure, 24 rue Lhomond,
              F-75231 Paris Cedex 05, France\\
              \email{patrick.hennebelle@ens.fr}
         \and
              Service d'Astrophysique, CEA/DSM/DAPNIA, C.E. Saclay,
              F-91191 Gif-sur-Yvette Cedex, France\\
              \email{pandre@cea.fr}
         \and
              AIM, Unit\'e Mixte de Recherche CEA -- CNRS -- Universit\'e
              Paris VII, UMR 7158
              }

   \date{Received December 27, 2005; accepted February 21, 2006}

 
  \abstract
   {The onset of gravitational collapse in cluster-forming clouds is still 
   poorly known.}
   {Our goal is to use the Class 0 protostar IRAS~4A, which is undergoing 
   collapse in the active molecular cloud NGC~1333, to set 
   constraints on this process. In particular we want to measure the mass
   infall rate and investigate whether the collapse could have been triggered
   by a strong external perturbation.}
   {We analyze existing continuum observations to derive the density structure
   of the envelope, and use our new molecular line observations done with the
   IRAM 30m telescope to probe its velocity structure. We perform a detailed
   comparison of this set of data with a numerical model of collapse triggered
   by a fast external compression.}
   {Both the density and velocity structures of the envelope can be well fitted
   by this model of collapse induced by a fast external compression for
   a time elapsed since point mass formation of 1-2 $\times$ 10$^{4}$ yr. We
   deduce a large mass infall rate of 0.7-2 $\times 10^{-4}$ 
   M$_\odot$~yr$^{-1}$. The momentum required for the perturbation to produce 
   this large mass infall rate is of the same order as the momenta measured 
   for the NGC~1333 numerous outflows. Our analysis shows also that the 
   turbulence is highly non uniform in the envelope, dropping from supersonic 
   to subsonic values toward the center. The inner subsonic turbulence is most
   likely a relic of the conditions prevailing in the dense core before the 
   onset of collapse.}
   {The vigorous collapse undergone by IRAS~4A was triggered by a fast external
   compression, probably related to the expansion of a nearby cavity, 
   which could have triggered the collapse of the 
   nearby Class 0 protostar IRAS~4B simultaneously. 
   This cavity could have been generated by an outflow but we have not found a 
   good protostellar candidate yet.}


   \maketitle
%

\section{Introduction}
\label{s:intro}

While the first phases of protostellar collapse in distributed star forming
regions like the Taurus molecular cloud start to be observationally better 
constrained \citep[e.g.][]{Tafalla98,Onishi99,Motte01,Belloche02}, little 
is still known about the onset of gravitational collapse in cluster-forming 
clouds where this process is likely to be more violent \citep[see][]{Andre04}.

The protostar IRAS~4A is located in the active molecular cloud NGC~1333 
forming low- and intermediate-mass stars in the Perseus complex 
\citep[e.g.][]{Sandell01}. The NGC~1333 region contains a double cluster of
infrared sources which is one of the least evolved embedded cluster known so
far, with an age of 1-2 $\times 10^6$ yr \citep[][]{Lada96,Lada03}. It contains
also a population of younger stellar objects, including a few Class 0 
protostars \citep[e.g.][]{Looney00}. IRAS~4A, 
one of these Class 0 protostars  \citep*[][]{Andre93},
harbors a 1.8$\arcsec$ binary system at a position angle P.A. of 
\mbox{-50$^\circ$} \citep[][]{Reipurth02} and is located in the vicinity of 
another young multiple system, IRAS~4B \citep[][]{Lay95}. 
Both sources are associated with molecular outflows oriented approximately in 
the North-South direction at 10$\arcsec$-scale \citep[e.g.][]{Choi01,Choi05}.
On larger scale, the IRAS~4A outflow has a P.A. $\sim 45^\circ$ 
\citep[e.g.][]{Blake95}. The whole region is actually filled 
with about ten molecular outflows driven by young protostars belonging to the 
NGC~1333 protocluster \citep[e.g.][]{Knee00}.

IRAS~4A was identified as a good infall candidate in the surveys of 
\citet{Mardones97} and \citet{Gregersen97}. \citet{DiFrancesco01} used the
IRAM Plateau de Bure interferometer to probe the inner parts of its envelope.
They detected inverse P-Cygni profiles in H$_2$CO(3$_{12}$-2$_{11}$) and 
CS(3-2) which they interpreted as infall motions. They derived a large mass
infall rate of $1.1 \times 10^{-4}$ M$_\odot$~yr$^{-1}$, which is about 70 
times larger than the standard accretion rate $\frac{c_s^3}{G}$ at 10 K 
\citep[][]{Shu77}. Such a large mass infall rate, if confirmed, can not occur 
in an envelope collapsing spontaneously. The collapse of the IRAS~4A envelope 
was therefore very likely triggered by a strong external perturbation. 
Several authors already argued that the numerous outflows 
have created cavities affecting the density structure of the molecular cloud 
and eventually triggered further star formation 
\citep[][]{Warin96,Lefloch98,Knee00,Quillen05}. In this respect, 
\citet{Sandell01} 
proposed that NGC~1333 is ``an example of self-regulated star formation''. 

However, it still remains to be shown that a collapse triggered by an external 
perturbation can match the density and velocity structure of a protostellar 
envelope such as IRAS~4A. This is the purpose of this work. The layout of
the paper is as follows. Sect.~\ref{s:obs} summarizes observational details.
In Sect.~\ref{s:results} we derive the density structure of the envelope from
a compilation of existing continuum data and we interpret in terms of 
velocity structure our new molecular line observations done with the IRAM~30m
telescope. We compare these results in Sect.~\ref{s:infall} to radiative
transfer models of hydrodynamical simulations of collapse triggered by a fast
increase of the external pressure \citep[][]{Hennebelle03,Hennebelle04}. 
Finally we discuss the implications in terms of triggered star formation in 
Sect.~\ref{s:disc}.


\section{Observations}
\label{s:obs}

We carried out milimeter line observations with the IRAM 30m telescope at Pico
Veleta, Spain, in September and October 2001, and in August 2004, in the 
following molecular transitions: HCO$^+$(1-0), H$^{13}$CO$^+$(1-0), 
HC$^{18}$O$^+$(1-0), N$_2$H$^+$(1-0), CS(2-1), C$^{34}$S(2-1), HCN(1-0) at 
3~mm, CS(3-2), C$^{34}$S(3-2), N$_2$D$^+$(2-1) at 2~mm, 
H$_2$CO(3$_{12}$-2$_{11}$), H$_2^{13}$CO(3$_{12}$-2$_{11}$), CS(5-4),  
N$_2$D$^+$(3-2) at 1.3~mm, and HCO$^+$(3-2), H$^{13}$CO$^+$(3-2) at 1~mm. The 
references to the frequencies we used are given in Sect.~\ref{ss:specsign}.
The half-power beamwidths can be computed with the equation HPBW (\arcsec) 
$= \frac{2460}{\nu(\mathrm{GHz})}$. We used four SIS heterodyne receivers
simultaneously and an autocorrelation spectrometer as backend. The spectral
resolution was 20 kHz at 3~mm and 2~mm, and 40 kHz at 1.3~mm and 1~mm in 2001,
and 10 kHz at 3~mm, and 20 kHz at 1.3~mm in 2004. The observations were done
in single-sideband mode with sideband rejections of 0.01 at 3~mm and 0.05 at
2, 1.3 and 1~mm. Accordingly, the calibration uncertainty was $\sim 10\%$.
The forward efficiencies $F_{\mathrm{eff}}$ were 0.95 at 3~mm, 0.93 at 2~mm, 
0.91 at 1.3~mm and 0.88 at 1~mm. The main beam efficiencies were computed 
using the Ruze function
$B_{\mathrm{eff}} = 1.2 \epsilon \,\, e^{-(4 \pi R \sigma / \lambda)^2}$, with
$\epsilon = 0.69$, $R \sigma = 0.07$ and $\lambda$ the wavelength in mm. The
system temperatures ranged from $\sim 110$ K to $\sim 160$ K at 3~mm, 
$\sim 190$ K to $\sim 250$ K at 2~mm, $\sim 320$ K to $\sim 670$ K at 1.3~mm 
and $\sim 550$ K to $\sim 1200$ K at 1~mm. The telescope pointing was checked
every $\sim$ 2 hours on Saturn, 3C84, and/or NRAO 140, and found to be 
accurate to 3 $\arcsec$ (rms). The telescope focus was optimized on Saturn 
and/or 3C84 every $\sim$ 3 hours. Position-switching and on-the-fly 
observations were done
with a reference position located at either (3000$\arcsec$,3000$\arcsec$),
(1500$\arcsec$,1500$\arcsec$), (0$\arcsec$,-240$\arcsec$), or 
(150$\arcsec$,120$\arcsec$) relative to the envelope center 
$\alpha_{\mathrm{J2000}} = 03^{\mathrm{h}}29^{\mathrm{m}}10.^{\mathrm{s}}30$,
$\delta_{\mathrm{J2000}} = 31^\circ13\arcmin31.8\arcsec$, as measured in
the 1.3 mm emission by \citet{Motte01}\footnote{Note that our central 
position is offset from the two components observed by \citet{Reipurth02} 
with VLA by 3$\arcsec$ and 1.5$\arcsec$ toward the West.}.
The data were reduced with the CLASS software package \citep[][]{Buisson02}.
The spectra were converted from antenna temperature to main beam temperature 
using the equation $T_{\mathrm{mb}}=\frac{F_{\mathrm{eff}}}{B_{\mathrm{eff}}} 
\, T_{\mathrm{a}}^*$.

In addition, we downloaded from the JCMT archive CS(5-4), CS(7-6), 
C$^{34}$S(5-4), HCO$^+$(4-3), H$^{13}$CO$^+$(3-2) and DCO$^+$(5-4) spectra 
observed toward IRAS~4A in February and July 1992, February 1994, 
November 1999, and October 2001. The data were converted into fits files with
SPECX \citep[][]{Padman93,Matthews97}, reduced with CLASS and converted to 
main beam temperatures using beam efficiencies $\eta_\mathrm{mb} = 0.69$ 
at 1~mm and 0.63 at 0.85~mm \citep[see][ for the conversion to 
$T_\mathrm{mb}$ at JCMT]{Maret04}.


\section{Results}
\label{s:results}

\subsection{Continuum emission}
\label{ss:cont}

\begin{table}
\centering
 \caption[]{Interferometric measurements of dust continuum emission.}
 \label{t:cont_interf}
 \begin{tabular}{lcccccl}
 \hline\hline
 Ref.$^{(1)}$ & \hspace*{-0.5cm} $\lambda$            & \hspace*{-0.3cm} HPBW                                     & \hspace*{-0.3cm} F$_{\mathrm{peak}}^{(2)}$ & \hspace*{-0.3cm} F$_{\mathrm{int}}^{(3)}$ & \hspace*{-0.3cm} Size     & \hspace*{-0.2cm} Interf. \\
                & \hspace*{-0.5cm} {\scriptsize (mm)} & \hspace*{-0.3cm} {\scriptsize ($\arcsec \times \arcsec$)} & \hspace*{-0.3cm} {\scriptsize (mJy.beam$^{-1}$)}     & \hspace*{-0.3cm} {\scriptsize (mJy)}                 & \hspace*{-0.3cm} {\scriptsize ($\arcsec \times \arcsec$)} & \\
 \hline
 D01  & \hspace*{-0.5cm} 1.33 & \hspace*{-0.3cm}  2.0 $\times$ 1.7  & \hspace*{-0.3cm} 1200 $\pm$ 16  & \hspace*{-0.3cm} 3100 $\pm$ 470 & \hspace*{-0.3cm} 5.3 $\times$ 4.1 & \hspace*{-0.2cm} PdBI \\
 D01  & \hspace*{-0.5cm} 2.04 & \hspace*{-0.3cm}  3.2 $\times$ 2.3  & \hspace*{-0.3cm} 450 $\pm$ 14   & \hspace*{-0.3cm} 690 $\pm$ 110  & \hspace*{-0.3cm} 4.4 $\times$ 3.4 & \hspace*{-0.2cm} NMA    \\
 D01  & \hspace*{-0.5cm} 2.19 & \hspace*{-0.3cm}  3.4 $\times$ 2.6  & \hspace*{-0.3cm} 420 $\pm$ 10   & \hspace*{-0.3cm} 580 $\pm$ 90   & \hspace*{-0.3cm} 4.2 $\times$ 3.3 & \hspace*{-0.2cm} NMA    \\
 L00  & \hspace*{-0.5cm} 2.7  & \hspace*{-0.3cm}  5.5 $\times$ 5.0  & \hspace*{-0.3cm} 351 $\pm$ 3    & \hspace*{-0.3cm} 544 $\pm$ 14   & \hspace*{-0.3cm} 25 $\times$ 24   & \hspace*{-0.2cm} BIMA \\
      & \hspace*{-0.5cm}      & \hspace*{-0.3cm}  3.0 $\times$ 2.8  & \hspace*{-0.3cm} 280 $\pm$ 2    & \hspace*{-0.3cm} 526 $\pm$ 9    & \hspace*{-0.3cm} 12 $\times$ 18.5 & \hspace*{-0.2cm} BIMA \\
      & \hspace*{-0.5cm}      & \hspace*{-0.3cm}  1.2 $\times$ 1.1  & \hspace*{-0.3cm} 172 $\pm$ 2    & \hspace*{-0.3cm} 450 $\pm$ 10   & \hspace*{-0.3cm} 5.4 $\times$ 6.2 & \hspace*{-0.2cm} BIMA \\
 D01  & \hspace*{-0.5cm} 3.22 & \hspace*{-0.3cm}  2.9 $\times$ 2.6  & \hspace*{-0.3cm} 200 $\pm$ 4.4  & \hspace*{-0.3cm} 320 $\pm$ 49   & \hspace*{-0.3cm} 4.5 $\times$ 3.7 & \hspace*{-0.2cm} PdBI   \\
 D01  & \hspace*{-0.5cm} 6.92 & \hspace*{-0.3cm}  2.1 $\times$ 1.9  & \hspace*{-0.3cm} 16  $\pm$ 0.29 & \hspace*{-0.3cm} 26 $\pm$ 4     & \hspace*{-0.3cm} 3.3 $\times$ 2.8 & \hspace*{-0.2cm} VLA    \\
 D01  & \hspace*{-0.5cm} 13.3 & \hspace*{-0.3cm}  3.8 $\times$ 3.1  & \hspace*{-0.3cm} 3.1 $\pm$ 0.27 & \hspace*{-0.3cm} 4.7 $\pm$ 1.0  & \hspace*{-0.3cm} 6.3 $\times$ 3.6 & \hspace*{-0.2cm} VLA    \\
 \hline
 \end{tabular}
 \begin{list}{}{}
 \item[$^{(1)}$] References: L00: \citet*{Looney00}; D01: \citet{DiFrancesco01}.
 \item[$^{(2)}$] Quoted errors are statistical errors only and do not include any systematic calibration uncertainties. 
 \item[$^{(3)}$] Flux integrated over a region of size listed in column 6.
 \end{list}
\end{table}

\begin{figure}[!t]
 \centerline{\resizebox{0.8\hsize}{!}{\includegraphics[angle=270]{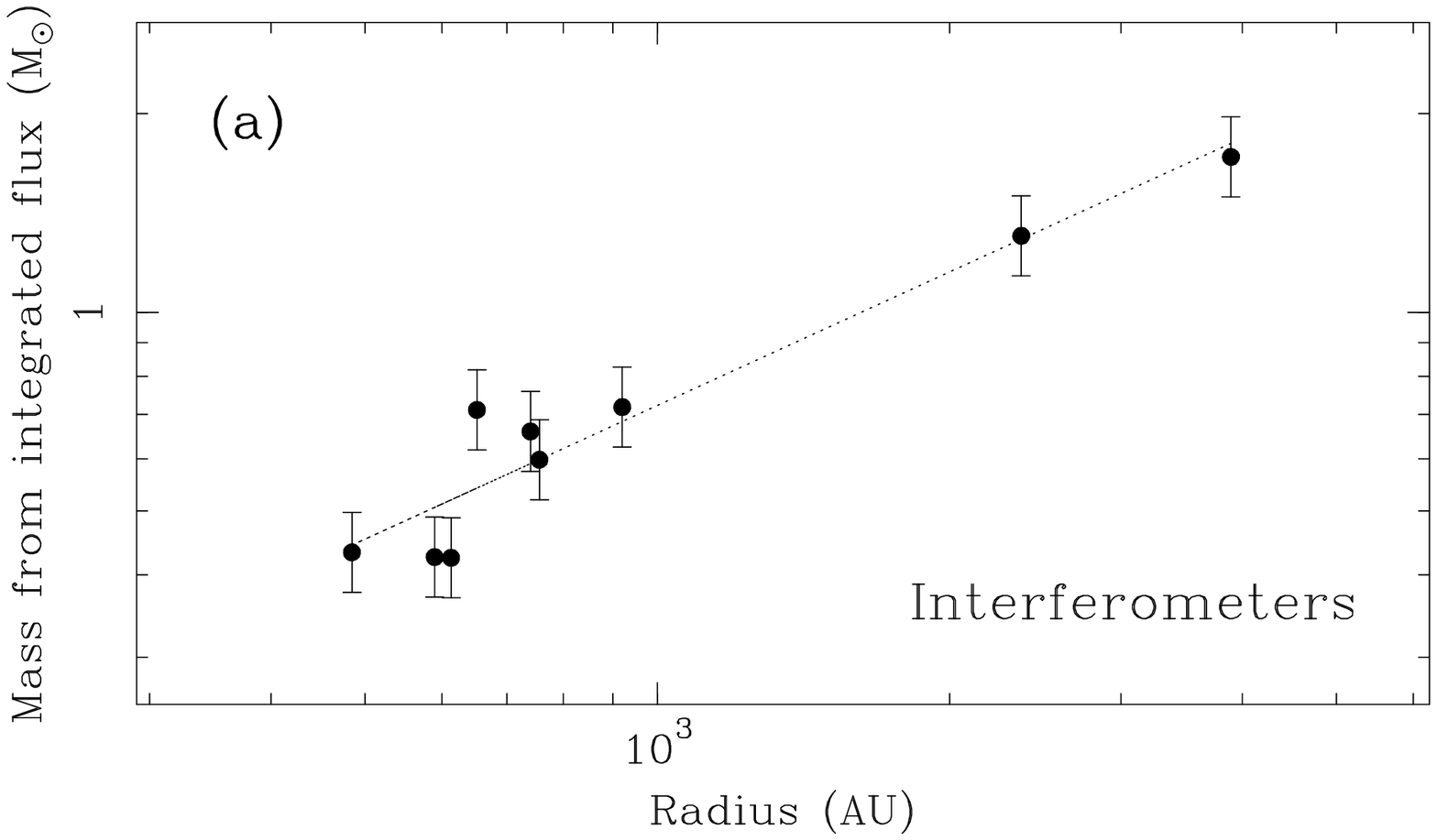}}}
 \centerline{\resizebox{0.8\hsize}{!}{\includegraphics[angle=270]{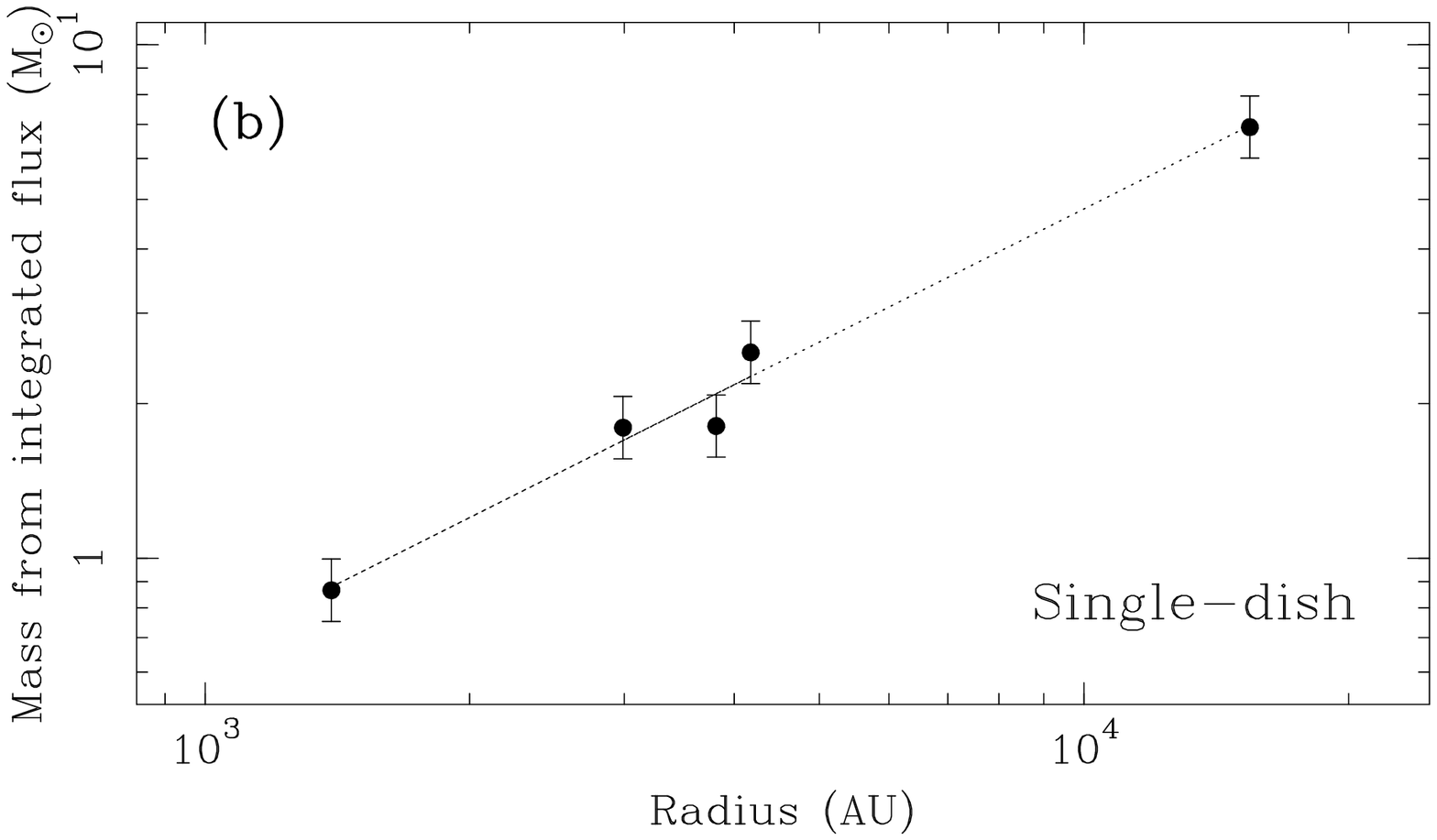}}}
 \vspace*{-1.0ex}
 \caption[]{Mass distribution estimated from the integrated flux of continuum 
 emission measured with \textbf{a)} interferometers (see 
 Tab.~\ref{t:cont_interf}) and \textbf{b)} single-dish telescopes (see 
 Tab.~\ref{t:cont_single}), and computed using the best-fit opacity exponent 
 $\beta = 0.85$.}
 \label{f:contmass}
\end{figure}

\begin{table}
\centering
 \caption[]{Single-dish measurements of dust continuum emission.}
 \label{t:cont_single}
 \begin{tabular}{lcccccl}
 \hline\hline
 Ref.$^{(1)}$ & \hspace*{-0.5cm} $\lambda$          & \hspace*{-0.3cm}  HPBW                     & \hspace*{-0.2cm} F$_{\mathrm{peak}}^{(2)}$ & \hspace*{-0.2cm} F$_{\mathrm{int}}^{(3)}$ & \hspace*{-0.2cm} Size                                     & \hspace*{-0.2cm} Telescope \\
              & \hspace*{-0.5cm} {\scriptsize (mm)} & \hspace*{-0.3cm} {\scriptsize ($\arcsec$)} & \hspace*{-0.2cm} {\scriptsize (Jy.beam$^{-1}$)}      & \hspace*{-0.2cm} {\scriptsize (Jy)}                 & \hspace*{-0.2cm} {\scriptsize ($\arcsec \times \arcsec$)} &                            \\
 \hline
 S00  & \hspace*{-0.5cm} 0.85 & \hspace*{-0.3cm} 16.0 & \hspace*{-0.2cm} 10.30 $\pm$ 0.03 & \hspace*{-0.2cm} 12.9 & \hspace*{-0.2cm} 30 $\times$ 23     & \hspace*{-0.2cm} SCUBA \\
 S01  & \hspace*{-0.5cm} 0.85 & \hspace*{-0.3cm} 14.0 & \hspace*{-0.2cm} ...              & \hspace*{-0.2cm} 9.05 & \hspace*{-0.2cm} 14.2 $\times$ 5.4  & \hspace*{-0.2cm} SCUBA \\
 L98  & \hspace*{-0.5cm} 1.25 & \hspace*{-0.3cm} 11.0 & \hspace*{-0.2cm} 3.10 $\pm$ 0.01  & \hspace*{-0.2cm} 4.7  & \hspace*{-0.2cm} 18.8 $\times$ 18.8   & \hspace*{-0.2cm} 30m   \\
 M01  & \hspace*{-0.5cm} 1.25 & \hspace*{-0.3cm} 11.5 & \hspace*{-0.2cm} 4.10 $\pm$ 0.02  & \hspace*{-0.2cm} 4.1  & \hspace*{-0.2cm} 24 $\times$ 24     & \hspace*{-0.2cm} 30m   \\
      &                       &                       &                                   & \hspace*{-0.2cm} 7.3  & \hspace*{-0.2cm} 97 $\times$ 97     & \hspace*{-0.2cm} 30m   \\
 \hline
 \end{tabular}
 \begin{list}{}{}
 \item[$^{(1)}$] References: L98: \citet{Lefloch98}; S00: \citet{Smith00}; S01: \citet{Sandell01}; M01: \citet{Motte01}.
 \item[$^{(2)}$] Quoted errors are statistical errors only and do not include any systematic calibration uncertainties. 
 \item[$^{(3)}$] Flux integrated over a region of size listed in column 6.
 \end{list}
\end{table}

The mass distribution of the IRAS~4A envelope can be estimated from the dust
continuum emission. We compiled from the literature the interferometric 
(Tab.~\ref{t:cont_interf}) and single-dish (Tab.~\ref{t:cont_single}) 
measurements done so far. Masses were estimated by various authors using 
different assumptions for the dust emissivity, the dust temperature 
distribution and/or the distance to NGC~1333. Here, we analyze this set of 
data homogeneously. We assume a distance of 318 pc, based on the 
\textit{Hipparcos} parallactic measurements of the Perseus OB2 association 
\citep[][]{deZeeuw99}, as discussed by \citet{Getman02}. With this distance,
the bolometric luminosity derived by \citet{Sandell91} becomes 11.6 L$_\odot$. 
We use a dust emissivity $\kappa_\lambda = 0.1 \times (\lambda/250 
\,\mu{\mbox{m}})^{-\beta}$ cm$^2$~g$^{-1}$ \citep[][]{Hildebrand83}. The dust 
temperature profile is assumed to be fixed by the heating by the central 
protostar as suggested by \citet{Terebey93}:
\begin{equation}
T_{\mathrm{d}}(r) = 25 \; \left(\frac{r}{6684 \; \mathrm{AU}}\right)^{-q} \; \left(\frac{L}{520 \; \mathrm{L}_\odot}\right)^{q/2},
\end{equation}
with $q=2/(4+\beta)$ and $L = 11.6$ L$_\odot$. We compute the mass within a 
spherical radius $R$ assuming optically thin emission and a temperature equal 
to $T_{\mathrm{d}}(R)$. However, this calculation overestimates the true mass 
because of the radius dependence of the temperature. When the temperature and 
the density are single power laws with slopes $-q$ and $-p$ respectively and
the Rayleigh-Jeans approximation is valid, the correction factor is 
$(1-\frac{q}{3-p})$. For radii smaller than $\sim 9200$ AU, 
$T_{\mathrm{d}}(r) > 10$ K and this approximation should be valid, provided
the density profile follows a power law. For larger radii, we expect the dust
grains to be also heated by the external UV radiation field via the 
photoelectric effect \citep[e.g.][]{Lesaffre05}. For simplicity, we assume 
that the dust temperature stays at 10 K for a radius larger than 9200 AU. In
this case, the correction factor to compute the mass within a radius $R > 9200$
AU is
\begin{equation}
 \left(\frac{q}{3-p-q} \times \left(\frac{R_1}{R}\right)^{3-p} + 1 
 \right)^{-1},
\end{equation}
where $R_1$ ($\sim 9200$ AU) is the radius where the temperature starts to be 
constant.

Given these assumptions, the emissivity exponent 
minimizing the mass dispersion of the interferometric measurements shown in 
Fig.~\ref{f:contmass}a is $\beta = 0.85 \pm 0.15$, yielding 
$\kappa_{\mathrm{1.3mm}} = 0.025 \pm 0.006$ cm$^2$~g$^{-1}$.
At larger radii with the 
single-dish measurements, we get $\beta = 1.1 \pm 0.6$, which is poorly 
constrained because these measurements are not spread out enough in 
wavelength. In the following, we use $\beta = 0.85$. The best power-law fits 
to the masses derived from the interferometric and single-dish measurements are
$M(r) = 1.9 \times (r/4200 \,\mbox{AU})^{0.67}$ M$_\odot$ and 
$M(r) = 3.1 \times (r/4200 \,\mbox{AU})^{0.86}$ M$_\odot$, respectively 
(see Fig.~\ref{f:contmass}). A fit to all the data gives $M(r) = 2.7 \times 
(r/4200 \,\mbox{AU})^{0.77}$ M$_\odot$. For comparison, the mass of the 
singular isothermal sphere enclosed in 4200 AU at 10 K is only 0.34 M$_\odot$.
We note that the interferometric 
measurements should resolve out any extended emission from the environment, 
and therefore give a more reliable estimate of the mass distribution within
13 $\arcsec$ in the IRAS~4A envelope.

Assuming a spherical geometry and a power-law density profile 
$n_{\mathrm{H}_2} = n_0 (\frac{r}{r_0})^{-p}$, the fit to the interferometric 
mass measurements yields $p = 2.3$ and $n_0 = 2.1 \times 10^5$~cm$^{-3}$ at 
$r_0 = 4200$~AU. By analyzing their 2.7 mm interferometric measurements, 
\citet{Looney03} found no evidence for the presence of a circumstellar disk 
and they measured a power-law index p between 1.7 and 2.3, which is consistent 
with our result (see their Tab.~1). Note also that $n_0$ is 4 times larger 
than the density of a singular isothermal sphere at 10 K, and this ratio is 
even larger at smaller radii.

\subsection{Spectroscopic signature of collapse}
\label{ss:specsign}

\begin{figure*}[!t]
 \centerline{\resizebox{0.30\hsize}{!}{\includegraphics[angle=0]{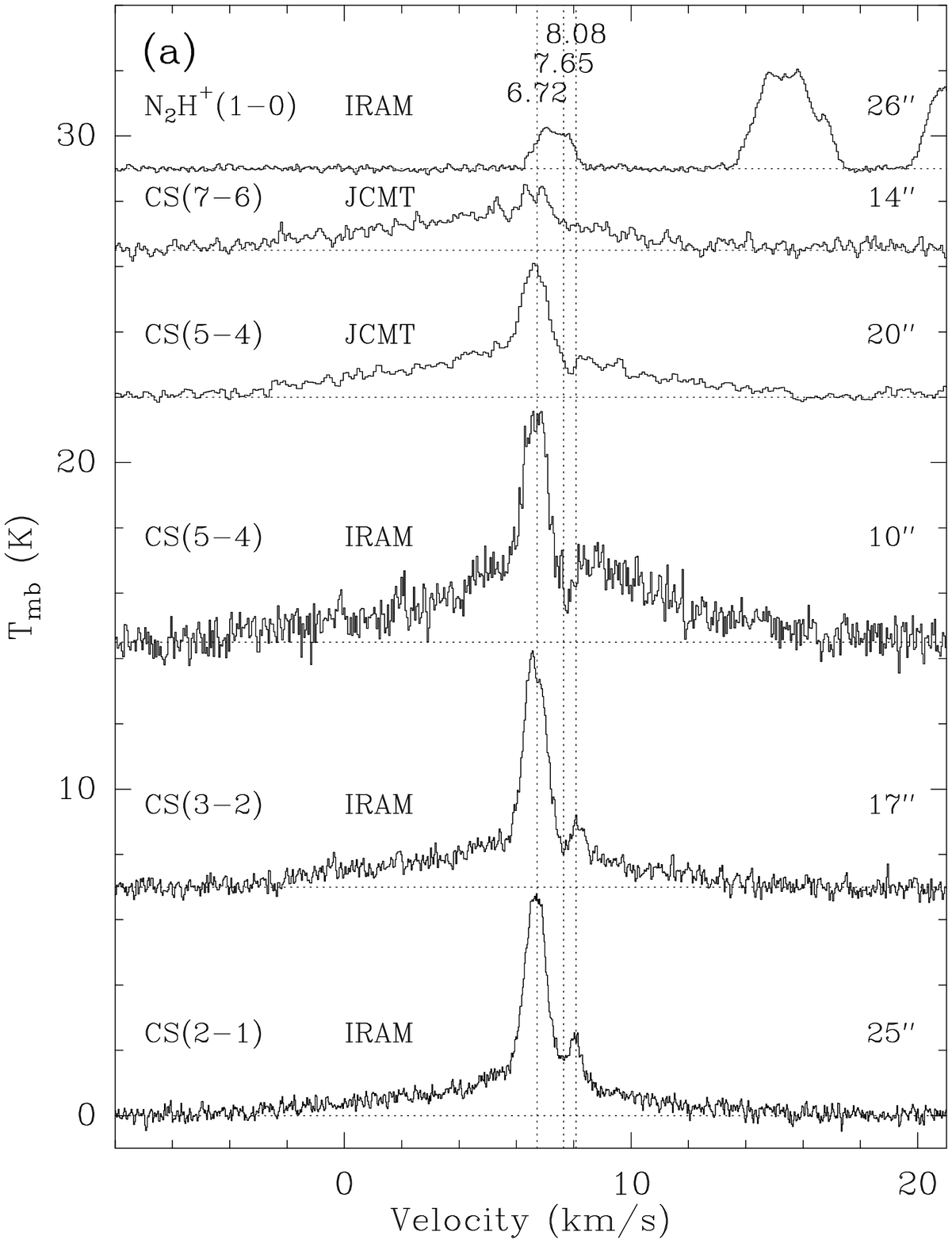}}\hspace*{0.03\hsize}\resizebox{0.30\hsize}{!}{\includegraphics[angle=0]{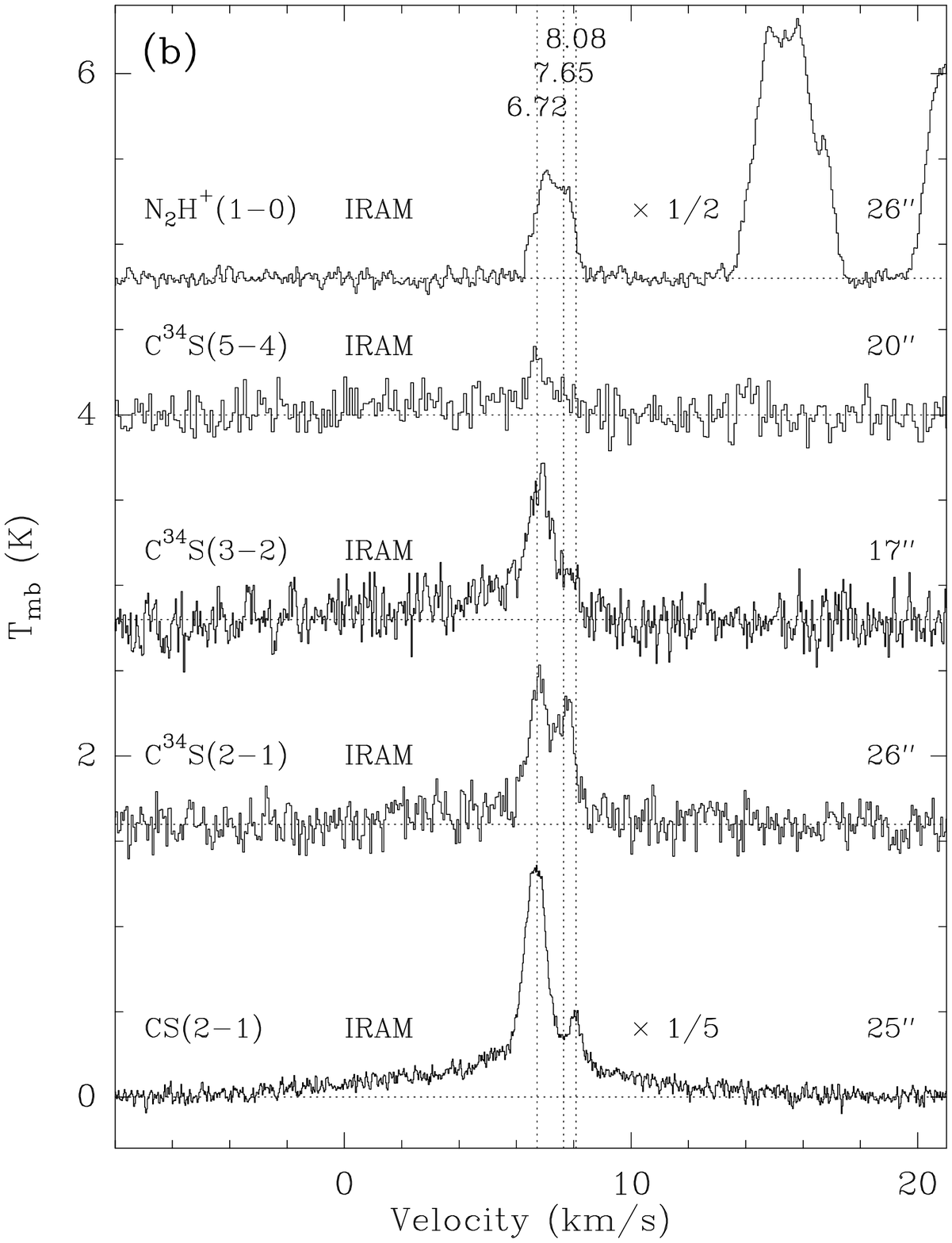}}\hspace*{0.03\hsize}\resizebox{0.30\hsize}{!}{\includegraphics[angle=0]{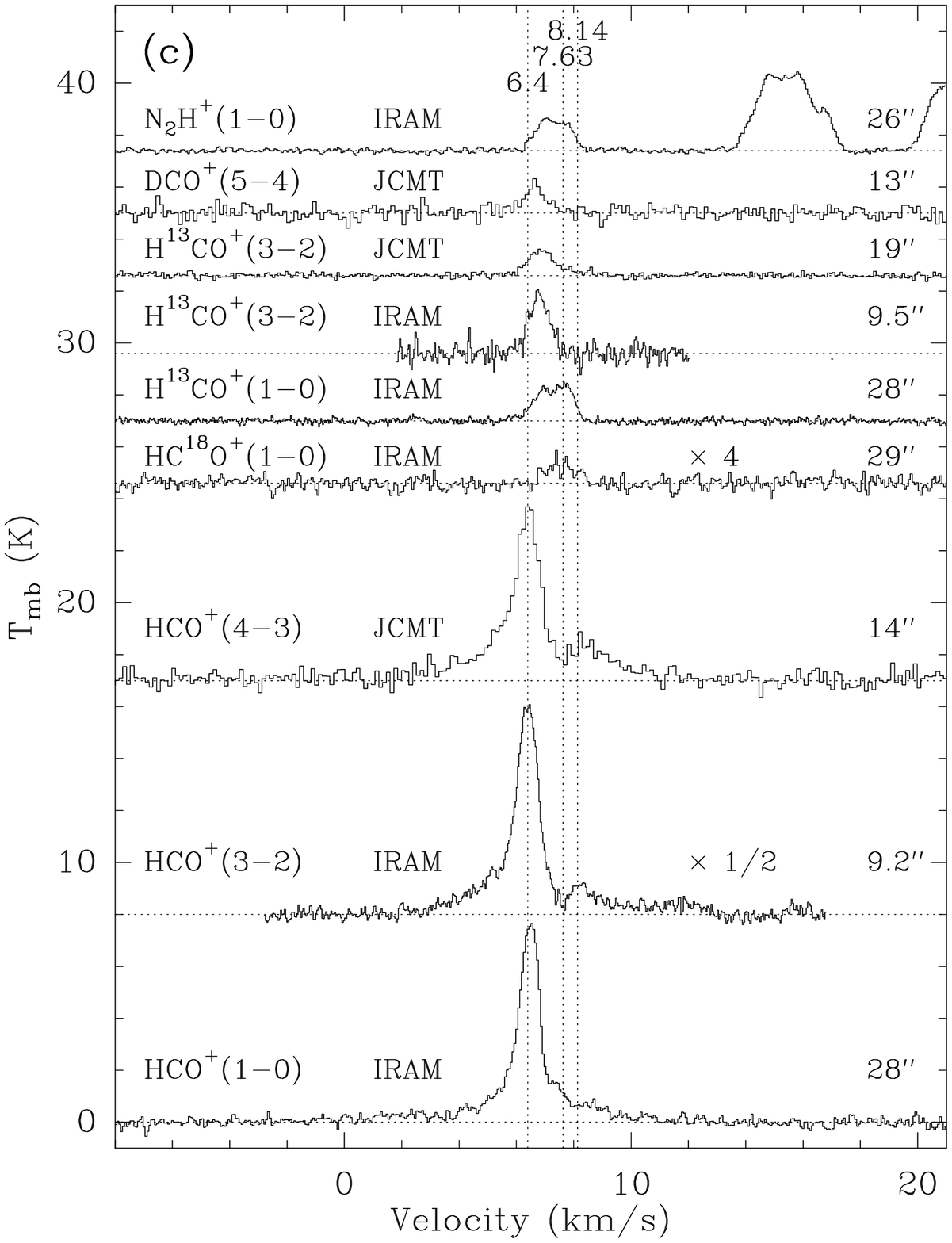}}}
 \vspace*{-1.0ex}
 \caption[]{Spectra toward the center of the IRAS~4A envelope obtained with 
 the 30m telescope or taken from the JCMT archive: \textbf{a)} CS and 
 N$_2$H$^+$ spectra, \textbf{b)} C$^{34}$S spectra along with CS(2-1) and 
 N$_2$H$^+$(1-0), \textbf{c)} HCO$^+$, H$^{13}$CO$^+$, HC$^{18}$O$^+$ and
 DCO$^+$ spectra along with N$_2$H$^+$(1-0).
 The vertical dotted lines show the position of the blue peak, the red peak 
 and the dip of the CS(2-1) spectrum in \textbf{a)} and \textbf{b)}, and the 
 same for HCO$^+$(3-2) in \textbf{c)}.
 The temperature scale is in T$_{\mathrm{mb}}$. The beamwidth (HPBW) 
 is given for each spectrum.}
 \label{f:centspec}
\end{figure*}

In Fig.~\ref{f:centspec} we compare the CS, C$^{34}$S, N$_2$H$^+$, HCO$^+$,
H$^{13}$CO$^+$ and HC$^{18}$O$^+$
spectra which we obtained toward the central position of IRAS~4A with the IRAM 
30m telescope, as well as CS, C$^{34}$S,  HCO$^+$, H$^{13}$CO$^+$ and DCO$^+$ 
spectra which we took from the JCMT archive. Among the latter, CS(5-4) and 
CS(7-6) were observed at a position slightly offset in right ascension (less 
than 3$\arcsec$). We used the CS and C$^{34}$S frequencies measured by 
\citet{Gottlieb03}, the N$_2$H$^+$(1-0) frequency determined by \citet{Dore04},
the DCO$^+$ frequencies measured by \citet{Caselli05}, the HCO$^+$ frequencies 
from the CDMS database \citep[as of October 2005,][]{Mueller05}, and the 
H$^{13}$CO$^+$ and HC$^{18}$O$^+$ frequencies determined by 
\citet{Schmid-Burgk04}.

The spectra of Fig.~\ref{f:centspec} show the classical signature of collapse, 
namely self-absorbed optically thick lines, with the blue peak being 
stronger than the red one, and low optical depth spectra peaking in between 
\citep[see][]{Evans99,Myers00}. This signature is seen at 20 $\arcsec$ in the 
East/West direction in our HCO$^+$(3-2) map, but is only seen on the central
position in our CS(5-4) map with a 10 $\arcsec$ spacing. Since the 
smoothing of our CS(5-4) map to the JCMT resolution yields a spectrum very 
similar to the JCMT one, the infall signature seen in the JCMT spectrum comes
entirely from the inner 10 $\arcsec$ probed by the IRAM 30m telescope. 

\begin{figure}[!t]
 \centerline{\resizebox{1.0\hsize}{!}{\includegraphics[angle=270]{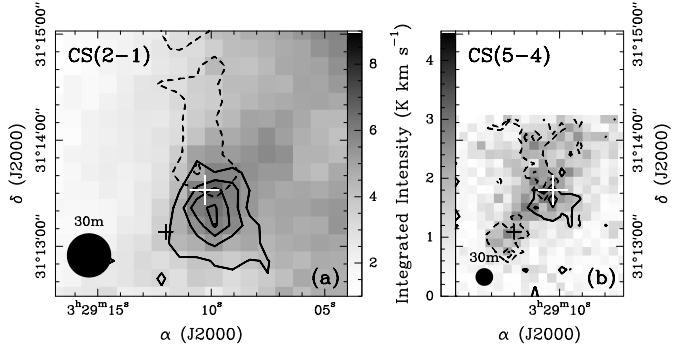}}}
 \vspace*{0.0ex}
 \caption[]{\textbf{a)} CS(2-1) intensity maps toward IRAS~4A integrated 
 over [-4.10,5.91] km~s$^{-1}$ (\textit{thick contours}), [6.06,8.31] 
 km~s$^{-1}$ (\textit{grey scale}), and [8.46,15.50] 
 km~s$^{-1}$ (\textit{dashed contours}). The contours go from 1.5 
 to 6 by  1.5 and from 1.5 to 3 by 1.5 K~km~s$^{-1}$, respectively.
 \textbf{b)} CS(5-4) intensity maps integrated over [-1.37,5.91] km~s$^{-1}$ 
 (\textit{thick contours}), [6.26,8.11] km~s$^{-1}$ (\textit{grey scale}),
 and [8.46,12.00] km~s$^{-1}$ 
 (\textit{dashed contours}). The contours are 2 and 4, and 
 0.7 and 1.4 K~km~s$^{-1}$, respectively. The intensity scale is 
 T$_{\mathrm{a}}^\star$. The IRAM 30m beam size is shown in each panel 
 (HPBW). The big/white and small/black crosses mark the positions of 
 IRAS~4A and IRAS~4B respectively.}
 \label{f:outflow}
\end{figure}

In the classical spectroscopic signature of collapse motions
\citep[][]{Evans99}, the shift of the dip of
the optically thick lines with respect to the systemic velocity of the source
given by the centroid velocity of an optically thin line is an indication of 
the infall velocity of the absorbing material. A fit to the hyperfine 
multiplet of N$_2$H$^+$(1-0) yields a low optical depth of $0.33 \pm 0.02$ for 
the isolated component 101-012, and a centroid velocity 
$V_{\mbox{\scriptsize N$_2$H$^+$}} = 7.24 \pm 0.01$ km.s$^{-1}$, which we take
as the systemic velocity of the source. With respect to this systemic velocity,
we measure shifts of 0.4, 0.4 and 0.5~km~s$^{-1}$ for CS(2-1), (3-2) and (5-4),
respectively, and shifts of 0.5 and 0.5~km~s$^{-1}$ for HCO$^+$(3-2) and (4-3),
respectively, all being measured with an uncertainty of 0.1~km~s$^{-1}$. 
\citet{DiFrancesco01} measured a self-absorption at 7.64 km~s$^{-1}$  with the 
Plateau de Bure interferometer in H$_2$CO(3$_{12}$-2$_{11}$), i.e. at the same 
velocity as the self-absorption in our CS(2-1) spectrum.

The C$^{34}$S(2-1) and (3-2) spectra are also asymmetric, with even a marked 
dip for C$^{34}$S(2-1). These lines could be optically thick and 
self-absorbed, like the CS lines. On the other hand, they could be optically 
thin but probe two velocity components along the line of sight. This would be 
the case for an infalling envelope with strong depletion at the center 
\citep[see the case of L1544 in][]{Caselli02}, or with a small 
non-thermal velocity dispersion compared to the infall velocities, or
for two independent velocity components, one being physically unrelated to 
the envelope. We will try to address this question in Sect.~\ref{s:infall} and 
\ref{s:disc}. With 
a signal-to-noise ratio of 3, only one peak is detected in the JCMT 
C$^{34}$S(5-4) spectrum, at the same velocity as the C$^{34}$S(2-1) blue 
component.

\subsection{Extended wings tracing the outflow}
\label{ss:wings}

In Fig.~\ref{f:outflow} we present contour maps of the CS(2-1) and CS(5-4) 
line wing emission overlaid on the intensity maps of the line core. Both
show a clear bipolar morphology whose direction P.A. $\sim 10^\circ$ matches
the direction of the outflow known in this source at small scale (see 
Sect.~\ref{s:intro}). The wings of the CS (and HCO$^+$) central spectra in 
Fig.~\ref{f:centspec} are thus certainly associated with the molecular 
outflow. They extend to larger velocities in CS (from $v = -2$ to 15 
km~s$^{-1}$) than in HCO$^+$ (3 to 11 km~s$^{-1}$).

\begin{figure}[!t]
 \centerline{\resizebox{1.0\hsize}{!}{\includegraphics[angle=270]{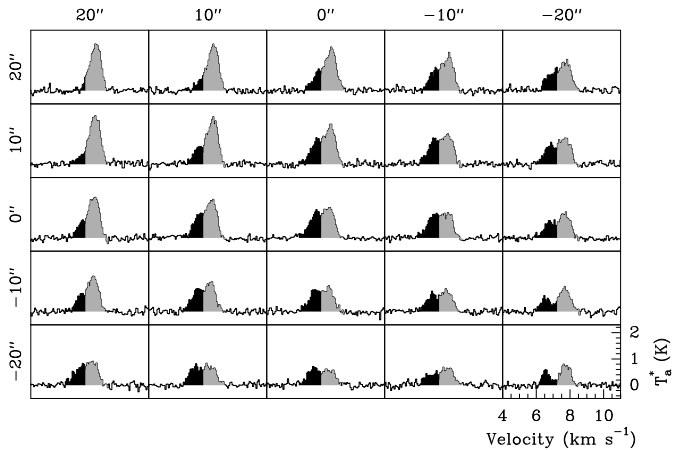}}}
 \vspace*{0.0ex}
 \caption[]{Map of H$^{13}$CO$^+$(1-0) spectra taken toward IRAS~4A with the 
 IRAM 30m telescope. Each spectrum is divided into two components, 
 filled in black for v $< 7.30$ km~s$^{-1}$ and in grey
 for v $ > 7.30$ km~s$^{-1}$. The half-power beam width is 28$\arcsec$.}
 \label{f:mapspec}
\end{figure}

\subsection{Maps of low optical depth tracers}
\label{ss:mapthin}

Figure \ref{f:mapspec} shows a map of H$^{13}$CO$^+$(1-0) spectra taken 
toward IRAS~4A with the IRAM~30m telescope. Nearly all spectra are 
double-peaked, as are also our C$^{34}$S spectra with a slightly worse 
signal-to-noise ratio. The intensity of the blueshifted component peaks at the 
IRAS~4A central position, this behaviour being more pronounced in the 
North-East/South-West direction than in the perpendicular direction. This 
component is thus certainly  physically 
associated with the IRAS~4A envelope. By fitting two Gaussian components to 
each spectrum, we measure for the blueshifted component a centroid velocity 
gradient of $\sim 9.7$ km~s$^{-1}$~pc$^{-1}$ with a 
position angle P.A. $= 38^\circ$ from North to East. 
On the other hand, the intensity of the redshifted component increases from the
South-West to the North-East, and shows only a small velocity gradient of 
2.4 km~s$^{-1}$~pc$^{-1}$ at P.A. $= 7^\circ$. The interpretation of
these velocity gradients in term of rotation is not straightforward.

\begin{figure}[!t]
 \centerline{\resizebox{1.0\hsize}{!}{\includegraphics[angle=270]{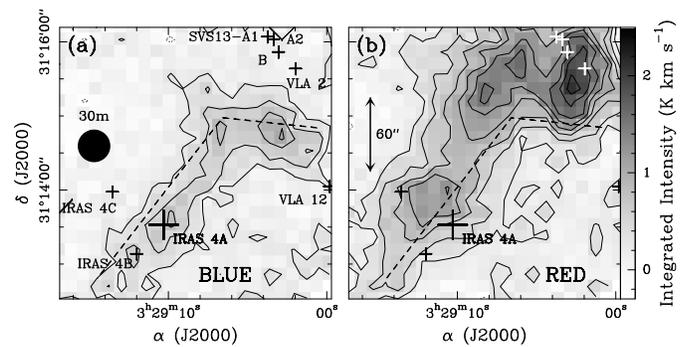}}}
 \vspace*{0.0ex}
 \caption[]{Maps of N$_2$H$^+$(101-012) intensity integrated over \textbf{a)} 
 the blueshifted velocity range [5.95,7.30] km~s$^{-1}$ and \textbf{b)} the 
 redshifted velocity range [7.30,8.65] km~s$^{-1}$, observed with the IRAM 30m 
 telescope (HPBW 26 $\arcsec$, shown on the left). 
 The first contour and contour step are 0.25 K~km~s$^{-1}$ in 
 T$_{\mathrm{a}}^\star$ scale in both maps.
 The big cross marks the central position which we used for IRAS~4A. The small 
 crosses mark the position of the VLA and BIMA continuum sources observed by 
 \citet{Reipurth02} and \citet*{Looney00}. The dashed line separates the two 
 components. It emphasizes the spatial shift between them and their shell-like 
 structure.}
 \label{f:mapn2hp}
\end{figure}

In Fig.~\ref{f:mapn2hp} we present large-scale maps of N$_2$H$^+$(101-012) 
intensity integrated over the same velocity ranges as for H$^{13}$CO$^+$(1-0) 
above (see Fig.~\ref{f:mapspec}). First of all, we notice that the 
N$_2$H$^+$(101-012) emission has a local maximum very close to IRAS~4A in both
Fig.~\ref{f:mapn2hp}a and b, which strongly suggests that \textit{both} 
the blueshifted and redshifted components are related to the protostar. 
Second, these maps show that the N$_2$H$^+$(101-012) emission traces the same 
structure as the dust continuum emission \citep[e.g.][]{Sandell01,Hatchell05}: 
a filamentary structure extending along a direction with a position angle P.A. 
$\sim$ -45$^\circ$, which bends at P.A. $\sim$ -90$^\circ$ in the West. 
Furthermore these maps also provide 
additional kinematical information and show that the two components which 
we identified in H$^{13}$CO$^+$(1-0) have the same morphology but are spatially
shifted: the ``blue'' filament is shifted by $\sim 30 \arcsec$ toward the 
South-West with respect to the ``red'' filament. Since they share the 
same morphology, they are likely to be physically related and not overlap by 
chance along the line of sight.

\section{Comparison to collapse models}
\label{s:infall}

The classical signature of infall seen in our CS and HCO$^+$ observations 
(see Sect.~\ref{ss:specsign} and Fig.~\ref{f:centspec}) suggests that 
the envelope of IRAS~4A is collapsing, confirming the analysis of 
\citet{DiFrancesco01}. The latter suggested that the inverse P-Cygni profiles 
which they observed with the IRAM Plateau de Bure Interferometer toward 
IRAS~4A in H$_2$CO(3$_{12}$-2$_{11}$) were created 
by an infalling envelope with a mass infall rate of $1.1 \times 10^{-4}$ 
M$_\odot$~yr$^{-1}$. This rate is nearly two orders of magnitude larger than 
the standard accretion rate $\frac{c_s^3}{G}$ at 10 K \citep[][]{Shu77}, and 
is the result of both larger densities and velocities (see 
Sect.~\ref{s:intro} and \ref{ss:cont}). It is thus very unlikely that the 
collapse of this protostellar envelope has occured spontaneously, and we 
investigate in this section models of collapse induced by a fast
external compression.

\subsection{Model of collapse induced by compression}
\label{ss:model}

\citet{Hennebelle03,Hennebelle04} investigated with a 
smoothed-particle-hydrodynamics code the collapse of prestellar and
protostellar cores driven from the outside by an increase of the external
pressure. In the following, we compare our observations to their non-rotating 
model with a rapid compression, i.e. $\phi =0.03$, where $\phi$ is the 
ratio of the time-scale on which the external pressure doubles to the initial 
sound-crossing time. Such a rapid compression is required to get
a density enhancement matching the density profile of the IRAS~4A envelope
(see Sect.~\ref{ss:model_mass}). On the other hand, a more rapid compression 
produces infall velocities too large to get a good agreement with our 
observations toward IRAS~4A. The initial conditions of the model correspond to 
a core in stable hydrostatic equilibrium. It is embedded in a hot rarefied 
external medium and has the structure of a Bonnor-Ebert sphere with a 
dimensionless radius $\xi_b = 3$, i.e. smaller than the dimensionless radius 
$\xi_b = 6.45$ of the critical Bonnor-Ebert sphere \citep[e.g.][]{Bonnor56}.
We normalized the model to a kinetic temperature of 10 K 
and a total mass of 6.0 M$_\odot$.

\subsection{Mass distribution}
\label{ss:model_mass}

\begin{figure}[!t]
 \centerline{\resizebox{1.0\hsize}{!}{\includegraphics[angle=270]{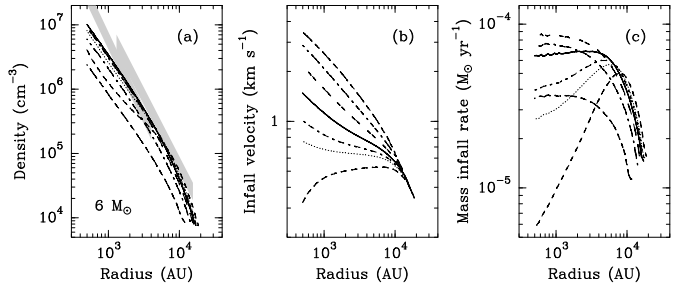}}}
 \vspace*{0.0ex}
 \caption[]{\textbf{a)} Density, \textbf{b)} infall velocity and \textbf{c)} 
 mass infall rate profiles of the collapse model induced
 by compression described in Sect.~\ref{ss:model} at times 0.8402, 0.8533, 
 0.8581, 0.8657, 0.8763, 0.8920 and 0.9249 $\times 10^6$ yr, corresponding to 
 renormalized times since the formation of the central protostar of 
 approximately -7, 7, 11, 19, 30, 45 and 78 $\times 10^3$ yr, respectively. 
 The thick line corresponds to 1.9  $\times 10^4$ yr and indicates 
 the ``best-fit'' model used in Sect.~\ref{ss:mapyso} and shown in  
 Fig.~\ref{f:profiles}.
 The dashed areas in (a) 
 show the constraints deduced for IRAS~4A from the interferometric and single 
 dish continuum measurements, respectively (see Sect.~\ref{ss:cont}).}
 \label{f:model}
\end{figure}

\begin{figure*}[!t]
 \centerline{\resizebox{1.0\hsize}{!}{\includegraphics[angle=270]{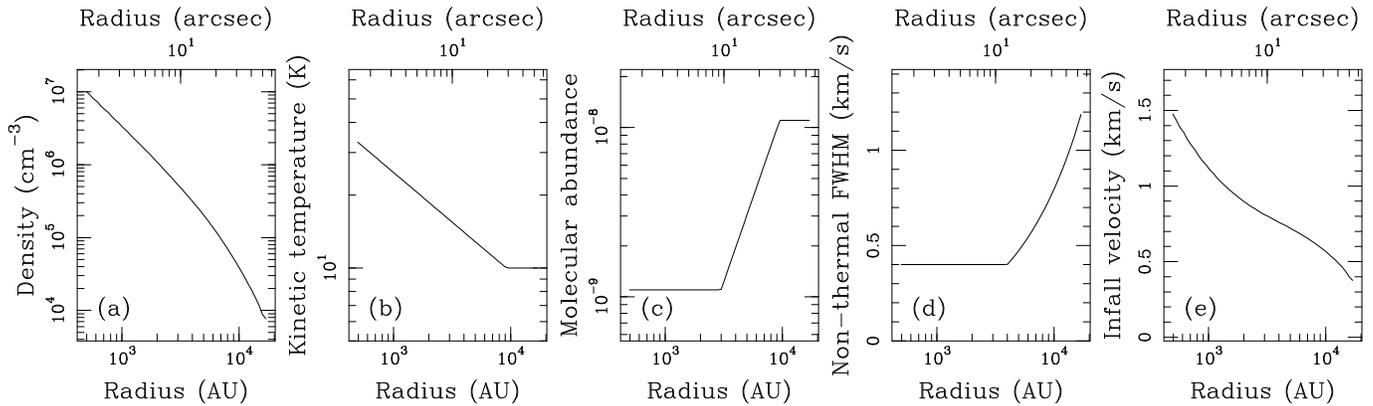}}}
 \vspace*{0.0ex}
 \caption[]{\textbf{a)} Density, \textbf{b)} kinetic temperature, \textbf{c)}
 CS molecular abundance, \textbf{d)} non-thermal line broadening (FWHM), and 
 \textbf{e)} infall velocity profiles of the collapse model at time
 $1.9 \times 10^4$ yr used for the radiative 
 transfer modeling of Fig.~\ref{f:compare}.} 
 \label{f:profiles}
\end{figure*}

Figure~\ref{f:model}a compares the evolution of the density profile of the 
model described in Sect.~\ref{ss:model} with the density profiles deduced from 
the interferometric and single-dish continuum measurements toward IRAS~4A in 
Sect.~\ref{ss:cont}. The fast external compression is responsible for 
a significant increase of the density profile with respect to the 
profile expected for a singular isothermal sphere. The agreement with the 
profiles deduced from the observations is good in the range 1500-10000 AU at 
times 1-3 $\times 10^4$ yr since point mass formation, which are of 
the same order as the estimated 
lifetime of Class 0 protostars \citep*[][]{Andre00}. The disagreement at 
smaller radii, the model being less dense than what we deduced from the 
observations assuming spherical symmetry, could be due to the presence of an 
unresolved disk not removed 
from the interferometric fluxes analyzed in Sect.~\ref{ss:cont}. However,
\citet{Looney03} did not find evidence for the presence of such a disk. At 
larger radii, the single-dish measurements might be contaminated by 
large-scale emission from the cloud along the line of sight, not related to 
the protostellar envelope itself, although the dual beam technique used for the
observations should have removed most of it.

\subsection{Radiative transfer modeling}
\label{ss:mapyso}

We used the Monte-Carlo-based radiative transfer code MAPYSO in one 
spherical dimension \citep[cf.][]{Blinder97,Belloche02}
to compute the CS, C$^{34}$S and N$_2$H$^+$ spectra that would be 
observed for a source
set in collapse by a fast compression, as described by the model of 
Sect.~\ref{ss:model}. In addition to the density and velocity profiles of the
hydrodynamical model at time 1.9 $\times 10^4$ yr (see Fig.~\ref{f:model} 
and \ref{f:profiles}a, e), we used a uniform CS/C$^{34}$S isotopic ratio of 
22.5 but non-uniform abundance profiles (see Fig.~\ref{f:profiles}c).  
For N$_2$H$^+$, we used the collision rates of HCO$^+$ with 
H$_2$ \citep[][]{Flower99} and assumed a constant abundance of 
$4.1 \times 10^{-10}$. We did not model the hyperfine structure. We simply
rescaled the abundance effectively used for the calculations by the 
statistical weight ratio $\frac{3}{27}$ and 
fitted only the 101-012 optically thin component. For all molecules,
we added a \textit{non-uniform} non-thermal line broadening to the thermal 
broadening (see Fig.~\ref{f:profiles}d).
Although the hydrodynamical 
simulation was isothermal, we used a non-uniform kinetic temperature to 
compute the spectra which are very sensitive to the heating by the central 
protostar. The temperature was set to the dust temperature profile described 
in Sect.~\ref{ss:cont} (see Fig.~\ref{f:profiles}b). 

\begin{figure}[!t]
 \centerline{\resizebox{1.0\hsize}{!}{\includegraphics[angle=270]{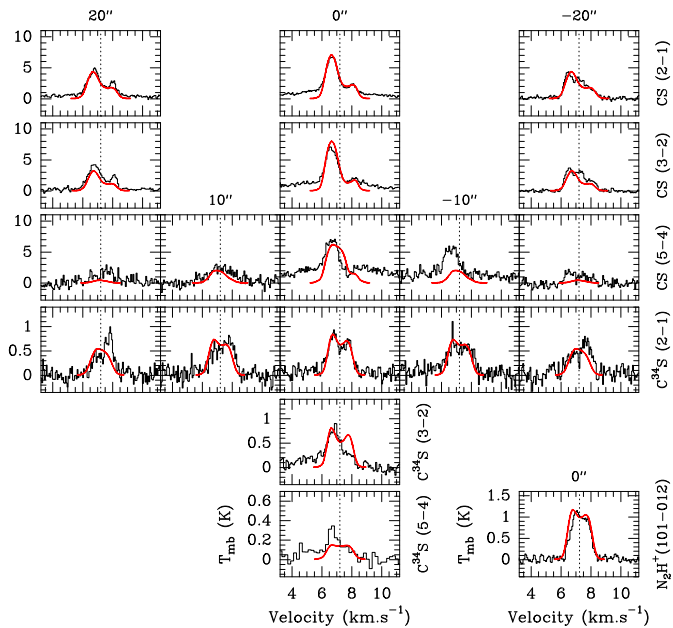}}}
 \vspace*{0.0ex}
 \caption[]{CS, C$^{34}$S and N$_2$H$^+$ spectra (histograms) in 
 units of main beam 
 temperature observed toward IRAS~4A along the longitude axis (see references
 in Fig.~\ref{f:centspec}). Synthetic spectra corresponding to the
 collapse model at time $1.9 \times 10^4$ yr described in 
 Sect.~\ref{ss:model} are superimposed in red/thick line. The dotted 
 line indicates the envelope systemic 
 velocity used for the computation of the radiative transfer 
 (7.20 km~s$^{-1}$ for CS and C$^{34}$S, and 7.24 km~s$^{-1}$ for 
 N$_2$H$^+$).}
 \label{f:compare}
\end{figure}

In Fig.~\ref{f:compare} we compare the synthetic spectra at time 1.9
$\times 10^4$ yr to the spectra observed toward IRAS~4A along the longitude 
axis, i.e. roughly perpendicular to the outflow axis (see 
Fig.~\ref{f:outflow}). The asymmetry and the absorption dips of the central 
CS(2-1) and CS(3-2) spectra are well reproduced. The asymmetry of these 
optically thick lines is enhanced toward the West and reduced toward the East 
at $\pm$ 20 $\arcsec$, which is not accounted for by our 1D spherical model. 
However including some rotation 
could improve the modeled spectra in that respect
\citep[see][ for such an effect in IRAM~04191]{Belloche02}. The line wings are
not reproduced by our model, which was expected since they trace the ouflow
that we did not include in our modeling (see Sect.~\ref{ss:wings}). The 
relative
contribution of the outflow gets stronger with the upper energy level of the
transition, and even masks the red peak of the CS(5-4) spectrum. This spectrum
is reasonably well matched by the model, although the shape of the dip is not 
fully reproduced.

The central C$^{34}$S(2-1) modeled spectrum matches well the observed one. It 
is optically thin ($\tau = 0.4$) and the two peaks result from the large 
infall velocities dominating the non-thermal line broadening in the inner 
parts of the envelope. We emphasize here the crucial importance of modeling
optically thin lines of a less abundant isotopologue \textit{simultaneously}
with the optically thick lines of the main molecule. This indeed reduces
the space of free parameters drastically -- especially the non-thermal line 
broadening and the abundance -- and helps to derive reliable infall velocities 
from the optically-thick, self-absorbed, asymmetric spectra. In particular, 
a first strong constraint is set on the abundance profile by the spatial 
variations of the C$^{34}$S(2-1) integrated intensity, which imply here a 
decrease of the abundance toward the center. Besides, a second constraint on
the abundance in the external parts comes from the depth of the absorption 
dips of the CS(2-1) and CS(3-2) lines. As a result, this set of data 
reveals that CS is strongly affected by depletion in the IRAS~4A enveloppe 
\citep[see][ for a similar result in the Class 0 protostar 
IRAM~04191]{Belloche02}.

The N$_2$H$^+$(101-012) line is also well reproduced, although the 
observed blue peak is not as blueshifted as in the model. The opacity of
the modeled spectrum is 0.24. It is in reasonable agreement with the opacity 
obtained from the Gaussian fit to the hyperfine structure 
(see Sect.~\ref{ss:specsign}), and the small discrepancy may be due to our 
simplified treatment of the radiative transfer of N$_2$H$^+$. 
With the uniform abundance, the N$_2$H$^+$(1-0) transition is more sensitive 
to the inner region than the C$^{34}$S(2-1) line. We tried a 
model including depletion of N$_2$H$^+$ above $\sim 10^{6}$ cm$^{-3}$ 
\citep[see][ for N$_2$H$^+$ depletion in IRAM~04191]{Belloche04} but the 
shape of the line was worse, the blue and red peaks being much more pronounced
than they are in the observed spectrum. The reason why the uniform abundance
produces a profile with less pronounced blue and red peaks is that in this 
case, the beam 
peaks up more material at small velocities projected along the line of 
sight, i.e. the material infalling along a direction close to the plane of the 
sky. This shows that the double-peaked profile of the C$^{34}$S(2-1) line is
not only due to the large infall velocities compared to the non-thermal 
broadening, as stated above, but also to the depletion in the central region. 
In our model, the shape of the N$_2$H$^+$(101-012) line therefore indicates no 
(or very little) depletion of N$_2$H$^+$ in the inner regions.

\begin{figure}[!t]
 \centerline{\resizebox{0.9\hsize}{!}{\includegraphics[angle=0]{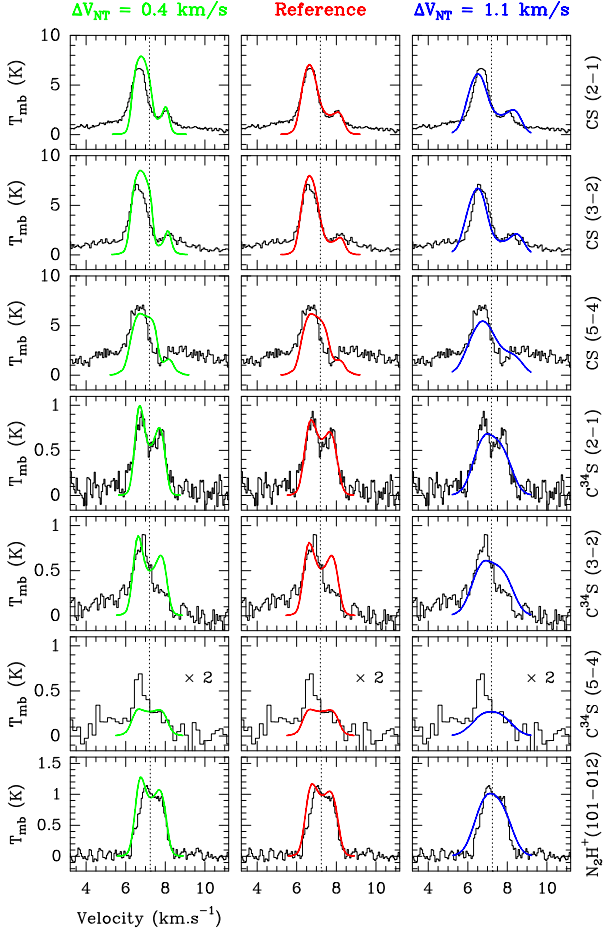}}}
 \vspace*{0.0ex}
 \caption[]{CS, C$^{34}$S and N$_2$H$^+$ spectra in units of main 
 beam temperature observed
 toward the center of IRAS~4A (\textit{histograms}). Each column shows
 a different model (\textit{thick line}). Synthetic spectra 
 corresponding to the collapse model at time $1.9 \times 10^4$ yr 
 described in Sect.~\ref{ss:model} are superimposed in the central column 
 (\textit{in red}). In the left and right columns, the 
 synthetic spectra  (\textit{in green and blue})
 were obtained with the same input model except for the non-thermal 
 broadening which was in both cases uniform, with $\Delta V_{\mathrm{NT}} = $ 
 0.4 and 1.1 km~s$^{-1}$, respectively. The dotted line indicates the envelope 
 systemic velocity used for the computation of the radiative transfer 
 (see Fig.~\ref{f:compare}).}
 \label{f:compare3}
\end{figure}

As we noted above, the non-thermal line broadening in the inner parts of the 
envelope is strongly constrained to be small ($\Delta V_{\mathrm{NT}} \sim 
0.4$ km~s$^{-1}$, i.e. $\sigma_{\mathrm{NT}} \sim 0.17$ km.s$^{-1}$) by the 
double-peaked line shape of the low-optical-depth C$^{34}$S(2-1) spectrum. The 
model with $\Delta V_{\mathrm{NT}} = 1.1$ km~s$^{-1}$ in 
Fig.~\ref{f:compare3} shows indeed that a larger broadening smoothes out the
C$^{34}$S(2-1) double-peaked line profile which traces the infall velocities 
of the front and rear hemispheres. It also broadens the 
N$_2$H$^+$(101-012) spectrum too much and produces wings which are not 
observed. On the 
other hand, the CS(2-1) and (3-2) broad absorption dips require a much larger 
non-thermal line broadening in the external parts of the envelope where they
are produced ($\Delta V_{\mathrm{NT}} \sim 1.2$ km~s$^{-1}$, i.e. 
$\sigma_{\mathrm{NT}} \sim 0.5$ km.s$^{-1}$). The model with 
$\Delta V_{\mathrm{NT}} = 0.4$ km~s$^{-1}$ in Fig.~\ref{f:compare3} shows 
indeed that a smaller broadening can not produce the broad absorption dips 
observed in CS(2-1) and CS(3-2). We have therefore a strong
evidence that \textit{the non-thermal line broadening is not uniform} in the 
IRAS~4A envelope, decreasing by nearly a factor of three toward the center, 
from supersonic down to
subsonic values. This conclusion agrees quite well with the small linewidths 
$\Delta V_{\mathrm{NT}} \sim$ 0.5 km~s$^{-1}$ observed toward IRAS~4A in 
N$_2$H$+$(1-0) by \citet{DiFrancesco01} at small scales with PdBI 
($r \sim 1000$ AU) and the much larger linewidths 
$\Delta V_{\mathrm{NT}} \sim$ 1.0-1.9 km~s$^{-1}$ measured in the NGC~1333 
molecular cloud with the low-density tracer C$^{18}$O(1-0) by \citet{Warin96}.

\begin{figure}[!t]
 \centerline{\resizebox{0.893\hsize}{!}{\includegraphics[angle=0]{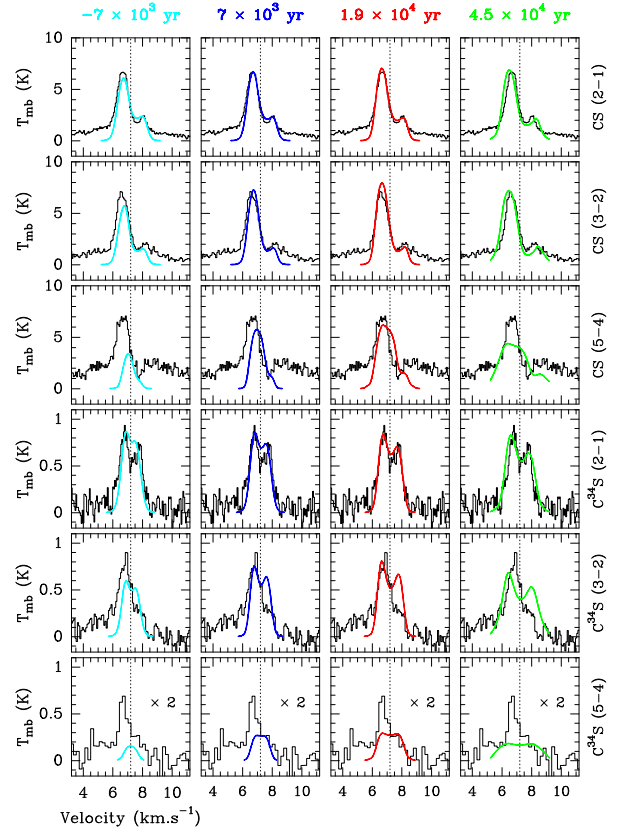}}}
 \vspace*{0.0ex}
 \caption[]{CS and C$^{34}$S spectra in units of main beam temperature observed
 toward the center of IRAS~4A (\textit{histograms}). Each column shows
 a different model (\textit{thick line}). Synthetic spectra 
 corresponding to the collapse model described in Sect.~\ref{ss:model} 
 at times -7, 7, 19 and 45 $\times 10^3$ yr are superimposed from left to 
 right (\textit{in light blue, blue, red and green}). 
 For each model, the scaling of the abundance profile was 
 adjusted to match the peak intensity of the C$^{34}$S(2-1) line.
 The dotted line indicates the envelope systemic velocity used for the 
 computation of the radiative transfer (see Fig.~\ref{f:compare}).
 }
 \label{f:compare2}
\end{figure}

The time elapsed since point mass formation in the collapse model
is well constrained by the peak separations and
the linewidths of the CS and C$^{34}$S spectra. The infall velocities of the 
model at time $7 \times 10^3$ yr in Fig.~\ref{f:compare2} are not large 
enough to produce the right separation of the peaks of the CS(3-2) and 
C$^{34}$S(2-1) lines. On the other hand, the much larger infall velocities of 
the model at time $4.5 \times 10^4$ yr produce too broad spectra. As a 
result, we find \textit{a good match for a time of 1-2 $\times 10^4$ 
yr elapsed since the formation of the central protostar}.

Finally, we could not find in the framework of our 1D spherical modeling a 
better match for the C$^{34}$S(3-2) and C$^{34}$S(5-4) central spectra. The 
modeled spectra of Fig.~\ref{f:compare} are not asymmetric enough for these 
two lines. Since our modeling suggests low optical depths of 0.35 and 0.04 for 
these two lines, it is very unlikely that the observed asymmetry results 
from an optical depth effect. This rather suggests that the material emitting 
in these two lines at the blue and redshifted velocities -- i.e. in the rear 
and front hemispheres, respectively, in the framework of this collapse model 
-- is denser for the former and less dense for the latter than assumed in the 
model.


\section{Discussion}
\label{s:disc}

\subsection{A vigorous collapse induced by a fast compression}

In Sect.~\ref{s:infall} we found a very good agreement between the continuum 
and CS/C$^{34}$S/N$_2$H$^+$ line observations toward IRAS~4A and the 
collapse model of 
\citet{Hennebelle03} induced by a fast external compression with 
$\phi = 0.03$. The best match is obtained for a time of 1-2 $\times 10^4$ yr 
elapsed since the formation of the central protostar. This corresponds to a 
mass infall rate of 4-7 $\times 10^{-5}$ M$_\odot$~yr$^{-1}$ in the range of 
radii [10$^3$,10$^4$] AU (see Fig.~\ref{f:model}c). However the actual mass 
infall rate is probably even larger since the density profile of our model 
does not match perfectly (and underestimates by about 25$\%$) the density 
profile deduced from the observations (see Fig.~\ref{f:model}a). Note also 
that the dust opacity we used in Sect.~\ref{ss:cont} is 2.5 times larger than 
the opacity usually assumed for protostellar envelopes \citep[e.g.][]{Motte01}.
Therefore we estimate that the actual mass infall rate is most probably in the 
range 0.7-2 $\times 10^{-4}$ M$_\odot$~yr$^{-1}$. This is in very good 
agreement with the mass infall rate of 1.1 $\times 10^{-4}$ 
M$_\odot$~yr$^{-1}$ deduced by \citet{DiFrancesco01} from their PdBI 
observations. As we already mentioned in Sect.~\ref{s:infall}, such a mass 
infall rate of 1 $\times 10^{-4}$ M$_\odot$~yr$^{-1}$ is 60 times larger than 
the standard accretion rate $\frac{c_s^3}{G}$ at 10 K \citep[][]{Shu77}. It is 
the result of both higher densities and velocities produced by the fast 
compression wave propagating inwards used in our model. 
\citet{Fatuzzo04} found indeed that the mass infall rate scales
roughly linearily with the initial overdensity and the initial velocity field 
in their self-similar collapse solutions where the dense core is initially 
either overdense compared to the singular isothermal sphere or has nonzero 
initial velocities. 

In the framework of the self-similar
inside-out collapse model with an effective sound speed including a 
non-thermal contribution, the ``standard'' accretion rate would be 
$\frac{(c_s^2+\sigma_{\mathrm{NT}}^2)^{3/2}}{G}$ \citep[][]{Shu87}. Using
the non-thermal velocity dispersion $\sigma_{\mathrm{NT}} = 0.17$ km~s$^{-1}$ 
derived in the inner parts of the envelope in Sect.~\ref{ss:mapyso}, which is
typical of the level of turbulence in prestellar condensations 
\citep[e.g.][]{Belloche01} but probably even an upper limit for IRAS~4A since 
the smallest dispersion measured by \citet{DiFrancesco01} with PdBI is 0.09 
km~s$^{-1}$, the ``effective'' accretion rate would be at most 3 
times larger than the pure isothermal one and cannot account for the large 
mass infall rate measured in the IRAS~4A envelope. 
On the other hand, the
Larson-Penston similarity solution \citep[][]{Larson69,Penston69} has a 
uniform mass infall rate of $29 \times \frac{c_s^3}{G}$ at point mass formation
and its extension after point mass formation by \citet{Hunter77} has a mass 
infall rate approaching asymptotically $47 \times \frac{c_s^3}{G}$. 
This value is closer to our result for IRAS~4A, which is not surprising since 
among the numerous similarity solutions of isothermal gravitational collapse 
the Larson-Penston solution can be physically interpreted as a model 
associated with a strong external compression wave \citep[][]{Whitworth85}.
However, although helpful to understand the physics of gravitational collapse, 
similarity solutions are not well suited for detailed comparison 
with observations since they do not have realistic boundary conditions.

By contrast to the large mass infall rate measured in the IRAS~4A 
envelope, the spontaneous collapse undergone by
the very young Class 0 protostar IRAM~04191 in the Taurus molecular cloud is
much less vigorous with a mass infall rate of $2 \times \frac{c_s^3}{G}$ only
\citep[see][]{Belloche02}. This could explain the order of 
magnitude difference between the bolometric luminosities of the two protostars 
which have yet approximately the same age since the beginning of the main 
accretion phase \citep[see][]{Lesaffre05}.

\subsection{Origin of the external compression}
\label{ss:origin}

The N$_2$H$^+$(1-0) maps of Fig.~\ref{f:mapn2hp} show that the material 
emitting at ``blueshifted'' velocities is displaced toward the South-West with 
respect to the material emitting at ``redshifted'' velocities and that both 
components have the same shell-like morphology, at the edge of cavity 2 of 
\citet{Lefloch98} seen in maps of continuum emission 
\citep[see also][]{Sandell01,Hatchell05}. We interprete the velocity 
difference between the two components as the result of ``inward'' motions and 
conclude that the ``blueshifted'' material is in the background and moving 
toward the ``redshifted'' material along a direction projected onto the plane 
of the sky at P.A. $\sim 45^\circ$ in the IRAS~4 region and P.A. 
$\sim 0^\circ$ in the SVS13/VLA~12 region. These geometry and velocity 
structure remind us of an expanding shell, the center of which would be 
located in the background and South-West of NGC~1333. The ``blueshifted'' 
component would be the expanding shell itself while the ``redshifted'' 
component would be the ambient medium. In this scenario, the whole region 
around IRAS~4 and IRAS~2 should show signatures of ``inward'' motion, i.e. the 
motion of the 
``blueshifted'' component toward the ``redshifted'' component. This might be 
indeed the case since \citet[][]{Walsh06} claim that they have detected 
``global infall'' in this region. On smaller scales in the IRAS~4A region, 
this scenario fits also, from a kinematical point of view, into the framework 
of the collapse model we investigated in Sect.~\ref{s:infall}. The external 
perturbation triggering the collapse of the dense core would then be the 
impact of this expanding shell on the ambient medium.

In this scenario, the origin of the external perturbation should be located in
the background and South-West of the IRAS~4/SVS13 region. The perturbation 
could result from the expansion of an H~II region, from a stellar wind or from 
a protostellar outflow. A nearby bright star in the South-West is required for 
the first two assumptions. The star BD +30$^\circ$547, with a spectral type 
between A 6 V and B 7 V 
\citep[][, although \citeauthor{Cernis90} 1990 found a type G 2 IV]{Aspin03}, 
would be the best candidate. However, \citet{Preibisch03} found with 
XMM-Newton that this star is in the foreground, so it cannot be, in our 
scenario, the origin of the external perturbation in the IRAS~4 region.
Since NGC~1333 is filled with many outflows 
\citep[][]{Knee00}, the collapse of IRAS~4 could have been triggered by one of 
them. Like \citet{Hennebelle03}, \citet{Motoyama03} investigated numerically 
the collapse of dense 
cores triggered by an external perturbation. They initiated the collapse by 
setting a nonzero inward velocity at the boundary of the dense core. They 
found that the maximum value of the mass infall rate is proportional to the 
momentum given to the dense core by the external perturbation, and in 
particular that an input momentum of $\sim 0.1$ M$_\odot$~km~s$^{-1}$ yields a 
mass infall rate of $\sim 1 \times 10^{-4}$ M$_\odot$~yr$^{-1}$. IRAS~4A could
thus have been set in collapse by an external perturbation with a momentum
of $\sim 0.1$ M$_\odot$~km~s$^{-1}$. Such a momentum is a typical value for
the present molecular outflows in NGC~1333 \citep[][]{Knee00}. The collapse
of IRAS~4A could therefore have been triggered, directly or indirectly, by the 
shock created by a present or former outflow. This would fit into the scenario 
of shock driven sequential star formation put forward by \citet{Warin96} in 
NGC~1333. Indeed, the cavity in the South-West of IRAS~4A could be the relic
of a former outflow \citep[see][]{Quillen05}. However, we have not yet
found a good protostellar candidate which could have been the origin of such 
an outflow in the South-West.

We mentioned the limitation of our 1D spherical modeling for IRAS~4A in 
Sect.~\ref{ss:mapyso} and suggested an asymmetry of the source along the line 
of sight, the rear hemisphere being denser than the front hemisphere. This 
asymmetry could result from the fast external compression itself, 
which is unlikely to be isotropic at the scale of the IRAS~4A envelope. 
\citet{Boss95} showed examples of such asymmetries in dense cores set
in collapse by a shock wave hitting only one hemisphere (see his Fig. 2). It
would be geometrically consistent with the idea that the compression wave 
which set the IRAS~4A envelope in collapse
is coming from the back, as deduced above from the morphology and kinematics 
of the N$_2$H$^+$(1-0) maps.

\subsection{Non uniform turbulence}
\label{ss:turbulence}

We found in Sect.~\ref{ss:mapyso} that the non-thermal line broadening, which
is likely to result from turbulent motions, decreases toward the center of 
IRAS~4A from supersonic to subsonic velocities by 
nearly a factor of 3. A 
similar conclusion was already drawn by \citet{DiFrancesco01} based on their 
N$_2$H$^+$ interferometric data. The turbulence is thus supersonic in the 
outer parts of the IRAS~4A envelope but only subsonic in the inner dense 
region. The subsonic turbulence in the inner part of the envelope
is reminiscent of the low level of turbulence found in prestellar condensations
\citep[e.g. in the $\rho$ Oph protocluster, see][]{Belloche01}. It could then 
indicate the conditions prevailing in the dense core before the onset of 
collapse. In the outer parts of the envelope, the supersonic turbulence 
$\sigma_{\mathrm{NT}} \sim 0.50$ km~s$^{-1}$ which we derived from the 
CS self-absorption is nearly as large as the amplitude of the non-thermal 
motions deduced by \citet{Warin96} from the low density tracer C$^{18}$O(1-0) 
in the molecular cloud \citep[see also Fig.~7 of][ for C$^{18}$O data with a 
better spatial resolution]{Quillen05}. The outer parts of the IRAS~4A envelope
are therefore affected by the supersonic turbulence which permeates the whole 
cloud and which was maybe powered by the numerous outflows located in NGC~1333
\citep[][]{Quillen05}. In the scenario we proposed in Sect.~\ref{ss:origin},
the supersonic turbulence could have been generated in the outer parts of the
IRAS~4A envelope by the expansion of the south-western cavity.

\subsection{Rotation and binary formation}

Detecting rotation and measuring its magnitude in the envelope of IRAS~4A is 
not straightforward. We mentioned in Sect.~\ref{ss:mapthin} that the blue 
component of the low optical depth tracers shows a velocity gradient of $\sim$ 
9.7 km~s$^{-1}$~pc$^{-1}$ at P.A. $\sim 38^\circ$ over $\sim 40\arcsec$, which 
is of the same order as the gradient measured in IRAM~04191 that we interpreted
as rotation \citep[][]{Belloche02}. However, there are two caveats for 
IRAS~4A. First, the double-peaked structure of our single-dish 
low-optical-depth spectra makes the analysis in terms of rotation difficult. 
Second, \citet[][]{DiFrancesco01} measured with PdBI in N$_2$H$^+$(1-0) 
centroid velocity differences of about 0.5-1 km~s$^{-1}$ over $\sim 12\arcsec$,
which corresponds to a velocity gradient of $\sim$ 30-50 km~s$^{-1}$~pc$^{-1}$,
at a position angle P.A. $\sim 135^\circ$. Their position angle was only a 
rough estimate and a fit to their centroid velocity map would actually give a 
position angle probably close to 90$^\circ$ (see their Fig.~2b), but this is 
still $\sim 50^\circ$ away from our single-dish measurement and it is unclear
whether these velocity gradients trace indeed rotation.

However, if we assume that the velocity gradient we measured in 
Sect.~\ref{ss:mapthin} is a good estimate of the amount of rotation in the 
envelope and suppose at first order that it is a solid-body rotation, then we 
find a ratio of rotation energy over gravitational energy of $\beta \sim 
0.02$. This is similar to the conditions investigated by \citet{Hennebelle04} 
to study the fragmentation process in a rotating core set in collapse by a 
fast external compression. Their main conclusion was that a fast 
compression 
promotes fragmentation and the formation of multiple protostars. With a very 
rapid compression ($\phi = 0.1$), they showed that this process 
occurs 
through the formation of a ring which becomes very quickly unstable. It breaks 
up into several pieces which, in the example they showed, merge into two 
protostars. For their simulation with a dense core of 1 M$_\odot$, the 
proto-binary separation is about 140 AU (see their Fig. 7), which corresponds 
to 840 AU if we normalize to the mass of the IRAS~4A envelope (6 M$_\odot$). 
IRAS~4A is indeed a binary system with a separation of 1.8$\arcsec$, i.e. 570 
AU at a distance of 318 pc \citep[see][]{Looney00,Reipurth02}. It is
therefore tempting to conclude that the formation of the IRAS~4A binary system
results from the fast external compression which has set up the dense 
core into collapse.

\subsection{Collapse of IRAS~4B}
\label{ss:iras4b}

Like toward IRAS~4A, \citet{DiFrancesco01} found an inverse P Cygni 
profile toward the Class 0 protostar IRAS~4B \citep[][]{Sandell91}.
They concluded that it is also collapsing, with a similar large mass 
infall rate of $3.7 \times 10^{-5}$ M$_\odot$~yr$^{-1}$. Since it lies along
the same N$_2$H$^+$ filament as IRAS~4A (see Fig.~\ref{f:mapn2hp}) and is in a 
similar evolutionary stage, we suggest that the gravitational collapse 
of its envelope was triggered by the same external perturbation, the expanding
shell suggested in Sect.~\ref{ss:origin}.

We note that IRAS 4B is an 11$\arcsec$ binary, and possibly a system 
with even higher multiplicity
\citep*[see][]{Lay95,Looney00}. However the main companion was not detected in 
continuum emission at 3.6cm by \citet{Reipurth02} and is probably much more 
evolved than the primary \citep*[][]{Looney00}. This raises the 
question whether it really belongs to the IRAS~4B system or is physically 
unrelated and appears just by chance projection close to the Class 0 
protostar.

\subsection{Alternative interpretations ?}

In the past, it has been claimed that the NGC~1333 molecular cloud includes 
several independant velocity components seen in emission 
and/or absorption \citep*[see][]{Langer96,Choi01,Choi04}. In particular, 
\citet{Choi04} rejected the infall interpretation of \citet{DiFrancesco01} and 
proposed instead a foreground absorbing layer at 8 km~s$^{-1}$ physically 
unrelated to IRAS~4A. They assumed a very low systemic velocity of 6.7 
km~s$^{-1}$ for IRAS~4A, yielding highly redshifted absorption dips in the 
optically thick spectra, and argued that standard collapse models like the 
inside-out collapse model \citep[][]{Shu77} would never produce such highly 
redshifted dips given the age of the protostar, and that the faster 
Larson-Penston model would produce much too broad 
linewidths in order to fit these highly redshifted dips. However, the model 
presented in Sect.~\ref{s:infall} shows that a systemic velocity of 
7.2 km~s$^{-1}$ and a collapse induced by a fast external compression can fit 
both the optically thick and thin lines reasonably well. The fit to our data 
set does not require any additional absorbing layer physically unrelated to 
IRAS~4A as claimed by \citet{Choi04}. Our scenario, which requires only one 
source, is simpler. Therefore we think it is more likely.


\section{Summary and conclusions}
\label{s:ccl}

We have carried out a detailed analysis of the physical structure of the
envelope of the Class 0 protostar IRAS~4A in the NGC~1333 molecular cloud. Our 
main results and conclusions are as follows:

\begin{enumerate}
\item The density profile deduced from previous interferometric and single-dish
millimeter continuum measurements is steep (p = 2.3) and the envelope is much 
denser than a singular isothermal sphere at 10 K.
\item Our new set of molecular line data obtained with the IRAM~30m telescope
shows the classical spectroscopic signature of collapse with absorption dips
of optically thick lines redshifted by 0.5 km~s$^{-1}$ with respect to the 
systemic velocity of the source traced by optically thin lines.
\item Both the density structure deduced from the continuum observations and 
our set of CS, C$^{34}$S and N$_2$H$^+$ spectra can be well fitted by 
a model of collapse induced by a fast external compression with a time elapsed 
since point mass formation of 1-2 $\times$ 10$^{4}$ yr. This age is 
consistent with the lifetime estimated for Class 0 protostars.
\item We deduce a very large mass infall rate of 0.7-2 $\times$ 10$^{-4}$ 
M$_\odot$~yr$^{-1}$ in the envelope of IRAS~4A. It results from both the 
overdensity and the large velocities generated by the fast external 
compression.
\item Our radiative transfer modeling reveals a strong decrease of the 
level of turbulence from supersonic velocities in the outer regions to 
subsonic velocities near the center of the IRAS~4A envelope. The subsonic 
turbulence is probably a relic of the conditions prevailing in the dense core 
before the onset of collapse.
\item We found indications of the presence of an expanding shell in the 
vicinity of IRAS~4A which could be the origin of the compression. 
This expanding shell could have triggered the collapse 
of the nearby Class 0 protostar IRAS~4B simultaneously.
However, we did not find evidence for the presence of a nearby bright 
star which could have initiated this compression through the expansion of an 
H~II region or a strong wind. On the other hand, the collapse of IRAS~4A 
and IRAS~4B could
have been triggered, directly or indirectly, by an outflow, but we have not
found a good protostellar candidate yet.
\end{enumerate}

\begin{acknowledgements}
We would like to thank Jennifer Hatchell for her help with the software SPECX
and Holger M\"uller for the new HCO$^+$ entry in the CDMS database. 
We thank also the referee, Mario Tafalla, for helpful comments.
\end{acknowledgements}

\end{document}